\documentclass[aps,prb,twocolumn,amsmath,amssymb,groupedaddress]{revtex4}

\usepackage{dcolumn}
\usepackage{graphicx,subfigure}
\usepackage{bm}
\usepackage{verbatim}
\usepackage{amsmath}
\usepackage{amssymb}
\usepackage[T1]{fontenc}
\usepackage{ae,aecompl}
\usepackage{appendix}

\newcommand{\rv}{{\bf r}}

\newcommand{\xv}{{\bf x}}
\newcommand{\kv}{{\bf k}}

\newcommand{\bv}{{\bf b}}

\newcommand{\qv}{{\bf q}}

\newcommand{\zh}{{\hat{\bf z}}}

\newcommand{\oh}{{\frac{1}{2}}}

\newcommand{\grad}{{\bm{\nabla}}}

\newcommand{\be}{\begin{equation}}
\newcommand{\ee}{\end{equation}}
\newcommand{\bea}{\begin{eqnarray}}
\newcommand{\eea}{\end{eqnarray}}
\newcommand{\bse}{\begin{subequations}}
\newcommand{\ese}{\end{subequations}}
\def\rf#1{(\ref{#1})}

\begin{document}
\title{Two-dimensional melting via sine-Gordon duality}
\author{Zhengzheng Zhai and Leo Radzihovsky}
\affiliation{
Department of Physics and Center for Theory of Quantum Matter\\
University of Colorado, Boulder, CO 80309}
\date{April 30, 2019}
\email{radzihov@colorado.edu}

\begin{abstract}
  Motivated by the recently developed duality [Pretko and Radzihovsky, Phys. Rev. Lett. {\bf 120}, 195301 (2018)] between the elasticity of a
  crystal and a symmetric tensor gauge theory, we
  explore its classical analog, which is a dual theory of the
  dislocation-mediated melting of a two-dimensional crystal,
  formulated in terms of a higher derivative vector sine-Gordon
  model. It provides a transparent description of the continuous
  two-stage melting in terms of the renormalization-group relevance of
  two cosine operators that control the sequential unbinding of
  dislocations and disclinations, respectively corresponding to the
  crystal-to-hexatic and hexatic-to-isotropic fluid transitions. This
  renormalization-group analysis reproduces the seminal results
  of the Coulomb gas description, such as the flows of the elastic
  couplings and of the dislocation and disclination fugacities, as
  well as the temperature dependence of the associated correlation
  lengths.
  
\end{abstract}
\pacs{}

\maketitle

\section{Introduction}
\subsection{Background and motivation}
The theory of continuous two-dimensional (2D) melting, developed by
Kosterlitz and Thouless\cite{Kosterlitz 72} , Halperin and
Nelson\cite{Nelson 79}, and Young\cite{Young 79} (KTHNY), building on
the work of Landau\cite{Landau 37}, Peierls\cite{Peierls 35} and
Berezinskii\cite{Berezinskii 71, Berezinskii 72}, has become one of
the pillars of theoretical physics.  Mathematically related to simpler
normal-to-superfluid and planar paramagnet-to-ferromagnet transitions
in films, described by a 2D XY model, it is a striking example of the
increased importance of thermal fluctuations in low-dimensional
systems\cite{Halperin 79, Nelson 83}. In contrast to their bulk
three-dimensional analogs, where, typically, fluctuations only lead to
{\em quantitative} modifications of mean-field predictions (e.g.,
values of critical exponents), here the effects are {\em qualitative}
and drastic.  Located exactly at the lower-critical dimension, a
local-order-parameter distinction between the high- and
low-temperature phases that is erased by fluctuations, two-dimensional
melting can proceed via a subtle, two-stage, {\em continuous}
transition, driven by the unbinding of topological dislocations and
disclinations defects. This mechanism, made possible by strong thermal
fluctuations, thus provides an alternative route to direct
first-order melting, argued by Landau's mean-field
analysis\cite{Landau 37} to be the {\em exclusive} scenario.

As such, the continuous two-dimensional melting (and related
disordering of a 2D XY model) is the earliest example of a
thermodynamically sharp, {\em topological} phase transition between
two locally disordered phases, which thus does not admit Landau's
local order-parameter description. It is controlled by a fixed line,
which lends itself to an asymptotically {\em exact}
analysis\cite{Kosterlitz 72, Nelson 79, Young 79}.

Although evidence for defects-driven phase transitions has appeared in
a number of experiments on liquid crystals\cite{Huang 93} and
Langmuir-Blodgett films\cite{Knobler 07}, finding simple model systems
that exhibit these phenomena in experiments or simulations has proven
to be more challenging. Most studied systems appear to exhibit 
discontinuous first-order melting that converts a crystal directly
into a liquid. However, it appears that two-stage continuous
melting has been experimentally observed by Murray\cite{Murray 88} and
Zahn\cite{Zahn 99} in beautiful melting experiments on two-dimensional
colloids confined between smooth glass plates and superparamagnetic
colloidal systems, respectively. In these experiments, an
orientationally quasi-long-range ordered but translationally
disordered hexatic phase\cite{Nelson 79} was indeed observed.  As was
first emphasized by Halperin and Nelson\cite{Nelson 79}, the hexatic
liquid, intermediate but thermodynamically distinct from the 2D
crystal and the isotropic liquid, is an important signature of the
defect-driven two-stage melting. In these two-dimensional colloids,
particle positions and the associated topological defects can be
directly imaged via digital video-microscopy, allowing precise
quantitative tests of the theory.

\subsection{Duality of the two-stage melting transition}
\label{sec:sketch}

The disordering of the simpler 2D XY model (describing, e.g., a
superfluid-normal transition in a film) is well known to admit two
complementary descriptions, the 2D Coulomb gas of
vortices\cite{CoulombGas} and its dual sine-Gordon field
theory\cite{sineGordon,sineGordonKT,Nelson 79, Halperin 79,
 ChaikinLubensky, RadzihovskyLubensky}. As with other dualities -- a subject with long
history and of much current interest\cite{Senthil 18} -- the
sine-Gordon duality has been extensively utilized in a variety of
physical contexts. Given that elasticity of a crystal can be thought
of as a space-spin coupled vector generalization of an XY model (with
vector phonon Goldstone modes $u_x,u_y$ replacing the scalar phase
angle), it is of interest also to develop an analogous dual
sine-Gordon formulation and to use it to study the 2D continuous
melting transition.

Indeed, recently, such a complementary description has emerged as a
classical limit of the elasticity-to-tensor gauge theory
duality\cite{PretkoRL 18/05, PretkoRL 18/12}, derived in the context
of a new class of topologically-ordered fracton
matter\cite{Nandkishore 18}.  As we will detail in the body of the
paper the corresponding dual Hamiltonian is given by
\begin{equation}
\begin{split}
  \tilde H &=\int d^2r \left[\frac{1}{2}
  \tilde{C}^{-1}_{ij,kl} \partial_i \partial_j
  \phi \partial_k\partial_l \phi\right.\\
&\left.-g_b \sum_{n=1}^{p} \cos({\bf b}_n
  \cdot \hat{{\bf z}} \times \grad \phi) 
-g_s \cos(s_p\phi)\right].
\end{split}
\end{equation}
Its key features, which characterize the continuous two-stage melting
are the higher-order ``Laplacian elasticity'', encoded via elastic
constants $\tilde{C}^{-1}_{ij,kl}$, and two sine-Gordon types of
operators with couplings $g_b, g_s$, respectively, capturing the
importance (fugacities) of dislocation (elementary vector charges
${\bf b}_n$) and disclination (elementary scalar charge $s_p$)
defects.

To flesh out the essence of this dual description, neglecting
inessential details, the above Hamiltonian is schematically described
by
\begin{equation}
  \tilde H \sim \int_\rv \left[ \oh \tilde C
 (\partial^2\phi)^2 - g_b\cos(\partial\phi) -
 g_s\cos(\phi)\right],
\end{equation}
where $\int_\rv\equiv\int d^2 r$.  Because of the second-order
Laplacian elasticity, standard analysis around the Gaussian fixed line
$g_b = g_s = 0$ shows that the mean-squared fluctuations of
Airy-stress potential $\phi$ diverge quadratically with system
size. This leads to an exponentially (as opposed to power-law in a
conventional sine-Gordon model) vanishing Debye-Waller factor, and in
turn to a strongly irrelevant disclination cosine, $g_s$, that can
therefore be neglected, whenever $g_b$ is small, i.e., near the
Gaussian fixed line.

\begin{figure}[htbp]
  \hspace{0in}\includegraphics*[width=0.5\textwidth]{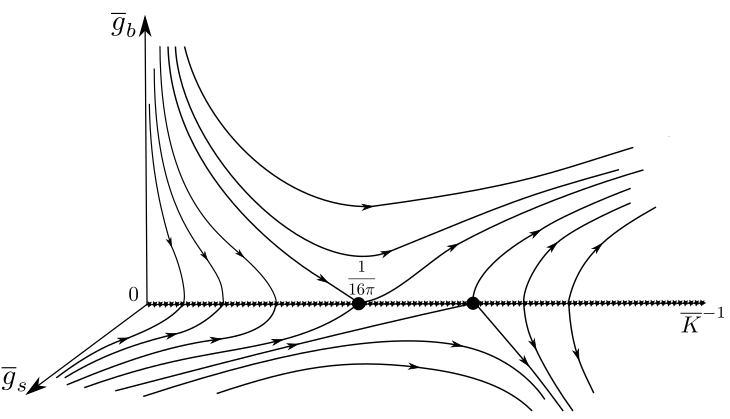}
  \caption{A schematic illustration of RG flows in the dual vector
    sine-Gordon model. It describes the two-stage continuous 2D
    melting, crystal-to-hexatic and hexatic-to-isotropic liquid
    transitions, associated with the consequent relevance of dislocations ($g_b$) and disclinations ($g_s$) fugacities, as a function of the elastic modulus ${\bar K}^{-1} =\frac{2\mu+\lambda}{4a^2\mu(\mu+\lambda)}$, expressed in terms of the Lam\'e elastic constants, $\mu,\lambda$ (defined in the main text) and the lattice constant $a$.}
\label{fig:RGflows}
\end{figure}

The schematic Hamiltonian then reduces to
\begin{equation}
    \tilde H_{\text{cr}} \sim \int_\rv
\left[ \oh \tilde C (\partial\chi)^2 - g_b\cos(\chi)\right],
\end{equation}
with $\chi = \partial\phi$. It thus obeys the standard sine-Gordon
phenomenology, exhibiting a KT-like ``roughenning'' transition in
$\chi$ with the relevance of $g_b$, controlled by the stiffness
$\tilde C$.\cite{sineGordon,sineGordonKT,Nelson 79, Halperin 79,
  ChaikinLubensky} At small $\tilde C < \tilde C_c$, $g_b$ is
irrelevant, describing the gapless crystal phase, with confined
dislocations and disclinations. The melting of the crystal is then
captured by the relevance of $g_b$ for $\tilde C > \tilde C_c$,
corresponding to a transition into a plasma of unbound dislocatons
characteristic of a hexatic fluid. Since in this phase $g_b$ is
relevant, at sufficiently long scales the dislocation cosine in
Eq. $(2)$ reduces to a harmonic potential for $\chi$,
$-g_b\cos(\partial\phi)\sim \oh g_b(\partial\phi)^2$. The effective
Hamiltonian is then given by
\begin{equation}
 \tilde H_{\text{hex}} \sim \int_\rv \left[\oh
  g_b(\partial\phi)^2 - g_s\cos(\phi)\right], 
\end{equation}
where we have neglected the $\tilde C$ ``curvature'' elasticity
relative to the gradient one encoded in $g_b$, and restored the
disclination cosine operator $g_s\cos(\phi)$. The resulting
conventional sine-Gordon model in $\phi$ can then exhibit the
second KT-like ``roughenning'' transition, capturing the
hexatic-to-isotropic fluid transition, asssociated with the unbinding of
disclinations. The corresponding RG flow of the dual vector
sine-Gordon model is schematically illustrated in
Fig. \ref{fig:RGflows}. We leave the detailed analysis of this two-stage
melting transition to the main body of the paper and the Appendix.
  
\subsection{Outline}
\label{sec:outline}

The rest of this paper is organized as follows. In Sec.
\ref{sec:Duality}, after briefly reviewing the elasticity theory of
two-dimensional crystal and its topological defects, we give two
detailed complementary derivations of the duality transformation to
the vector sine-Gordon model. Utilizing the latter we
straightforwardly reproduce known results for the crystal-hexatic
phase transition in Sec.  \ref{sec:RGsineGordon}, by focusing on the
dislocation fugacity cosine operator and neglecting the irrelevant
disclinations. Inside the hexatic phase, we derive a scalar
sine-Gordon model for the Airy stress potential, that captures the
subsequent hexatic-isotropic liquid transition.  We conclude in
Sec. \ref{sec:summary} with a summary of our results and a discussion of
potential applications of this dual approach.


\section{Duality of 2D melting}
\label{sec:Duality}
\subsection{ Two-dimensional elasticity}
\label{sec:elasticity}
At low temperatures, the deformations of a crystal do not vary
substantially over the atom size, allowing it to be described with a
continuum field theory of its phonon Goldstone modes, $u({\bf r})$,
with a short-distance cutoff set by the lattice spacing $a$. The
underlying translational symmetry (spontaneously broken by the
crystal), requires that the elastic energy is an analytic expansion in
the strain field $\partial_i u_j$. Due to the underlying rotational symmetry
in the target space (i.e., no substrates and/or external fields), the
elastic Hamiltonian is further constrained (in harmonic order) to be
independent of the antisymmetric part of $\partial_i u_j$, i.e., of
the local bond angle $\theta(\rv)=\frac{1}{2}
\epsilon_{ij} \partial_iu_j=\frac{1}{2}\hat{\bf z}\cdot\grad\times{\bf
  u}$. The elastic Hamiltonian density to harmonic order is thus given
by
\begin{equation}
    \mathcal{H}=\frac{1}{2} C_{ij,kl} u_{ij}u_{kl},
 \label{H_el}
\end{equation}
where $u_{ij}=\frac{1}{2}(\partial_i {\bf r} \cdot \partial_j {\bf
  r}-\delta_{ij})$ is the symmetric nonlinear strain tensor (fully
rotationally invariant in the target space), which in the harmonic
approximation takes the linear symmetrized form
\begin{equation}
    u_{ij} \approx \frac{1}{2} (\partial_i u_j+\partial_j u_i).
\label{harmonic_u}
\end{equation}
$C_{ij,kl}$ is the elastic constant tensor, whose number of
independent components is restricted by the symmetry of the
crystal. For simplicity, we focus on the isotropic hexagonal lattice,
where $C_{ij,kl}$ takes the form
\begin{equation}
    C_{ij,kl}=\lambda \delta_{ij}\delta_{kl}+ 2\mu\delta_{ik}\delta_{jl}
\end{equation}
and characterized by two independent elastic constants, namely the Lam\'e
coefficients, $\lambda$ and $\mu$. As we discuss in Appendix
\ref{appendixB}, an external stress $\sigma^e_{ij}(\rv)$ is included
through an additional term $-\sigma^e_{ij} u_{ij}$, here focusing on
the case of a vanishing external stress.

\subsection{ Topological defects}
\label{sec:defects}
In addition to the single-valued elastic phonon fields, the crystal
also exhibits topological defects -- disclinations and dislocations,
captured by including a nonsingle-valued part of the phonon distortion
field.

\begin{figure}[htbp]
  \centering
  \subfigure[]{
    \includegraphics[width=1.8in]{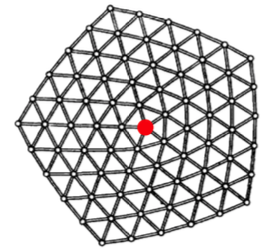}}
    \label{fig:disclination}
  \hspace{1.0cm}
  \subfigure[]{
    \includegraphics[width=2.0in]{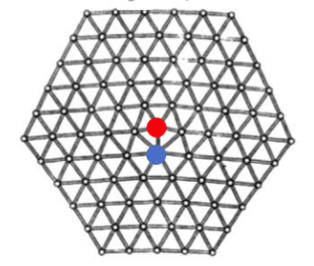}}
    \label{fig:dislocation}
    \caption{Topological defects in a 2D hexagonal lattice.  (a) A
      disclination. (b) A dislocation; a dipole of two, opposite
      charge disclinations. (Figure adapted from Ref. [20].)}
\label{fig:defects}
\end{figure}

Disclinations are topological defects associated with orientational
order.  A disclination at a point ${\bf r}_0$, illustrated in
Fig. \ref{fig:defects}(a), is defined by a nonzero closed
line-integral of the gradient of the bond angle around ${\bf r}_0$:
\begin{equation}
   \oint_{\rv_0} d\theta=\frac{2\pi }{p} s
 \end{equation}
 or equivalently in a differential form:
\begin{equation}
    \hat{\bf
      z}\cdot\grad\times\grad\theta=\frac{2\pi}{p}s\delta^2({\bf
      r}-\rv_0)\equiv \frac{2\pi}{p}s({\bf r}),
\label{defn_s}
\end{equation}
measuring the deficit/surplus bond angle, $(2\pi/p)s$, with $s$ the
integer disclination charge in a $C_p$ symmetric crystal. In the case
of a hexagonal lattice, $p=6$. In the above equation, $s({\bf r})$ is the disclination charge density.

Dislocations are vector topological defects associated with
translational order. A dislocation at ${\bf r}_0$ with a Burgers
vector-charge ${\bf b}_n$ (that is an elementary lattice vector), as
illustrated in Fig. \ref{fig:defects}(b), is defined by a closed
line-integral
\begin{equation}
    \oint_{\rv_0} d{\bf u}= {\bf b}_n,
\end{equation}
or equivalently in the differential form,
\begin{equation}
  \hat{\bf z}\cdot\grad\times\grad u_i=b_{i,n}\delta^2({\bf
    r}-\rv_0)\equiv b_i({\bf r}),
\label{defn_b}
\end{equation}
where ${\bf b}({\bf r})$ is the Burgers charge density.  As
illustrated in Fig. \ref{fig:defects}(b), a dislocation is a
disclination dipole, and it is therefore energetically less costly than a
bare disclination charge.

\subsection{ Vector Coulomb gas formulation}
\label{sec:CoulombGas}
In the presence of topological defects, the distortion field ${\bf
  u}({\bf r})$ is not single-valued. Along with the associated strain
tensor, the distortion field can be decomposed into the single-valued elastic phonon and the
singular parts,
\begin{eqnarray}
u_i&=&\tilde{u}_i+ u_i^s,\\
u_{ij}&=&\tilde{u}_{ij}+ u_{ij}^s.
\end{eqnarray}
To include these topological and phonon degrees of freedom we focus on
the partition function (taking $k_B T = 1$, i.e., measuring coupling
constants in units of thermal energy),
\begin{equation}
\begin{split}
Z&= \int [d{\bf u}]e^{-\int_\rv \mathcal{H}[{\bf u}]}\\
&=\int [d{\bf\tilde u}][d{\bf u}^s]\int [d\sigma_{ij}({\bf r})]
e^{-\int_\rv\mathcal{H}[{\bf u}, \sigma_{ij}]},
\end{split}
\end{equation}
where the trace over ${\bf u}(\rv)$ implicitly includes both phonons
and topological defects by allowing nonsingle-valued distortions. In
the second form, above, we decoupled the elastic energy by introducing
a Hubbard-Stratonovich tensor field -- the {\em symmetric} stress
tensor $\sigma_{ij}(\rv)$\cite{ChaikinLubensky}, with the resulting
Hamiltonian density given by
\begin{equation}
\begin{split}
  \mathcal{H}[{\bf u},\sigma_{ij}] &=\frac{1}{2} C^{-1}_{ij,kl}
  \sigma_{ij} \sigma_{kl}+i\sigma_{ij}u_{ij}\\
  &=\frac{1}{2} C^{-1}_{ij,kl} \sigma_{ij}\sigma_{kl}+i
  \sigma_{ij}\left( \partial_i \tilde{u}_j + u_{ij}^s \right).
\end{split}
\end{equation}
Above, for a 2D hexagonal lattice, 
\begin{equation}
C^{-1}_{ij,kl}=-\frac{\lambda}{4\mu (\mu+\lambda)} \delta_{ij}
\delta_{kl}+\frac{1}{2\mu}\delta_{ik}\delta_{jl}
\label{Cinv}
\end{equation}

Tracing over the single-valued phonons $\tilde{\bf u}$, enforces the
divergenless stress constraint [via $\delta$-function identity $\frac{1}{2\pi}\int_{-\infty}^{\infty} du  e^{iuf} =\delta(f)$]
\begin{equation}
\partial_i \sigma_{ij}=0,
\end{equation}
solved with a scalar Airy stress potential, $\phi(\rv)$,
\begin{equation}
    \sigma_{ij}=\epsilon_{ik}\epsilon_{jl} \partial_k\partial_l\phi.
\end{equation}
Expressing the Hamiltonian density in terms of $\phi(\rv)$, and 
integrating by parts in the second linear term, we utilize the defects
conditions, Eq. (\ref{defn_s}) and (\ref{defn_b}),
\begin{subequations}
\begin{eqnarray}
\epsilon_{ik} \epsilon_{jl} \partial_l \partial_k u_{ij}^s&=&
\epsilon_{ik} \epsilon_{jl} \partial_l \partial_k(\partial_i u_j^s
-\epsilon_{ij}\theta^s),\\
&=&\epsilon_{ki}\partial_k b_i(\rv) +
\epsilon_{ki}\partial_k\partial_i\theta(\rv),\\
&=&\zh\cdot\grad\times{\bf b} + \frac{2\pi}{6}s(\rv),
\end{eqnarray}
\end{subequations}
to obtain
\begin{eqnarray}
\label{Hphi}
\mathcal{H}[\phi]
&=&\frac{1}{2} \tilde{C}_{ij,kl}^{-1} \partial_i\partial_j
\phi \partial_k\partial_l \phi
+ i\phi \left(\frac{2\pi}{6} s 
+ \zh\cdot\grad\times{\bf b}\right).\nonumber\\
\end{eqnarray}
Above $\tilde{C}^{-1}_{ij,kl}=\epsilon_{ia} \epsilon_{jb}
\epsilon_{kc} \epsilon_{ld} C^{-1}_{ab,cd}$.


Focussing on dislocations and neglecting the high energy disclination
defects, we can straightforwardly integrate out $\phi(\rv)$ in the
partition function, obtaining a dislocations vector Coulomb gas
Hamiltonian 
\begin{equation}
H_{\bf b}=\frac{1}{2} \int \frac{d^2 q}{(2\pi)^2}  b_i({\bf q})
\tilde K_{ij}({\bf q}) b_j(-{\bf q}),
\end{equation}
where the tensor Coulomb interaction in Fourier and coordinate spaces
is given by
\begin{eqnarray}
\tilde K_{ij}({\bf q})
&=&\frac{K}{q^2}\left(\delta_{ij}- \frac{q_iq_j}{q^2}\right),\\ 
K_{ij} ({\bf r})&=&-\frac{K}{4\pi} 
\left(\delta_{ij}\ln(r/a) -\frac{r_i r_j}{r^2}\right),
\label{Kij}
\end{eqnarray}
with $K=\frac{4\mu(\mu+\lambda)}{2\mu+\lambda}$.

Thus, dislocations vector Coulomb gas Hamiltonian in real space
reduces to,
\begin{equation}
\begin{split}
   H_{\bf b}&=-\frac{K}{8\pi} \int_{\rv_1,\rv_2}
   \left[{\bf b}({\bf r}_1) \cdot {\bf b}({\bf r}_2) \ln \frac{|{\bf
         r}_1-{\bf r}_2|}{a} \right. \\
& \left.-\frac{{\bf b}({\bf r}_1) \cdot ({\bf r}_1-{\bf r}_2) 
({\bf r}_1-{\bf r}_2)\cdot{\bf b}({\bf r}_2)}{|{\bf r}_1-{\bf r}_2|^2}
\right],
\end{split}
\end{equation}
which, in the discrete form and supplemented with core energies (see
below) is exactly the vector Coulomb gas model used by Nelson and
Halperin\cite{Nelson 79} and by Young\cite{Young 79}, as the theory of 2D continuous two-stage melting.

\subsection{Dual vector sine-Gordon model}
Motivated by the sine-Gordon description of the XY model, we dualize
elasticity by transforming the above vector Coulomb gas into a vector sine-Gordon model and re-examine the two-stage continuous 2D melting transition from this complementary approach.

Dislocation and disclination densities on a hexagonal lattice are
given as a sum of their discrete charges
\begin{eqnarray}
{\bf b}({\bf r})
&=&\sum_{{\bf r}_n}{\bf b}_{{\bf r}_n}\delta^2({\bf r}-{\bf r}_n),\\
s({\bf r})
&=&\sum_{{\bf r}_n} s_{{\bf r}_n} \delta^2({\bf r}-{\bf r}_n),
\end{eqnarray}
where ${\bf r}_n=a(n_1 \hat{\bf e}_1+n_2 \hat{\bf e}_2)$, ($n_1, n_2
\in \mathbb{Z}$) are triangular lattice vectors spanned by unit
vectors $\hat{\bf e}_1= \hat{\bf x}$ and $\hat{\bf
  e}_2=\frac{1}{2}\hat{\bf x}+\frac{\sqrt{3}}{2} \hat{\bf y}$, and ${\bf b}_{{\bf r}_n}=a(n_1 \hat{\bf e}_1+n_2 \hat{\bf e}_2)$, and $s_{{\bf r}_n} \in \mathbb{Z}$ are dislocation and disclination charges, respectively.

In terms of these discrete topological defect charges, the Hamiltonian
is given by
\begin{equation}
\begin{aligned}
  H=& \frac{1}{2} \int_{\rv}\,
  \tilde{C}^{-1}_{ij,kl} \partial_i \partial_j
  \phi \partial_k \partial_l \phi+\sum_{{\bf r}_n} \left[\tilde E_{b}
    {\bf
      b}^2_{{\bf r}_n} +E_{s} s^2_{{\bf r}_n}\right]\\
  & +\sum_{{\bf r}_n} \left[i\frac{2\pi}{6} \phi({\bf r}_n)
    s_{{\bf r}_n} - i \hat{\bf z} \times \grad \phi({\bf
      r}_n) \cdot {\bf b}_{{\bf r}_n}\right],
\label{Hdiscrete}
\end{aligned}
\end{equation}
where we have added dislocation and disclination core energies $E_{b}=
a^2\tilde E_{b}$ and $E_{s}$ to account for the defects' energetics at
the scales of the lattice constant, not accounted for by the
elasticity theory\cite{Nelson 79}. The partition function involves an
integration over potential $\phi(\rv_n)$ and summation over the
dislocation and disclination charges. Following a standard
analysis\cite{Kosterlitz 72,Nelson
  79,CoulombGas,sineGordon,ChaikinLubensky},
\begin{widetext}
\begin{equation}
\begin{split}
    Z=&\int [d\phi] \sum_{\{s_{{\bf r}_n}\}} 
\sum_{\{{\bf b}_{{\bf r}_n}\}} \prod_{{\bf r}_n} 
e^{-H[\phi, {{\bf b}_{{\bf r}_n}},{s_{{\bf r}_n}}]} ,\\
=&\int[d\phi]
e^{-\oh \int_{\rv} \tilde{C}_{ij,kl}^{-1} \partial_i \partial_j
  \phi \partial_k \partial_l \phi}
\sum_{\{s_{\rv_n}\}}\sum_{\{\bv_{\rv_n}\}}
\prod_{\rv_n}\left[e^{-i\frac{2\pi}{6}\phi(\rv_n) s_{\rv_n}-E_{s}
    s_{\rv_n}^2} 
e^{i\zh\times\grad\phi(\rv_n)\cdot\bv_{\rv_n}-\tilde E_{b}b_{\rv_n}^2}\right],\\
=&\int[d\phi]
e^{-\oh \int_{\rv} \tilde{C}_{ij,kl}^{-1} \partial_i \partial_j
  \phi \partial_k \partial_l \phi}
\left[1+e^{-2E_{s}}\int \frac{d^2 r_1}{a^2} \frac{d^2 r_2}{a^2}
e^{-i\frac{2\pi}{6}\phi(\rv_1)}e^{i\frac{2\pi}{6}\phi(\rv_2)} +
\ldots\right]\\
&\left[1+e^{-2E_{b} }\sum_{n=1}^3 
\int\frac{d^2 r_1}{a^2} \frac{d^2 r_2}{a^2}
e^{-i\zh\times \grad\phi(\rv_1)\cdot\bv_n}
e^{i\zh\times \grad\phi(\rv_2)\cdot\bv_n}+\ldots\right],\\
=&\int[d\phi]e^{-\oh \int_{\rv}
  \tilde{C}_{ij,kl}^{-1} \partial_i \partial_j \phi 
\partial_k \partial_l \phi} \cdot 
\left[1+e^{-E_{s}}\int \frac{d^2 r_1}{a^2}
\left(e^{i\frac{2\pi}{6}\phi(\rv_1)}+e^{-i\frac{2\pi}{6}\phi(\rv_1)}\right)
\right.\\
& \left.
+ \frac{1}{2!}e^{-2E_{s}}\int \frac{d^2 r_1}{a^2} \frac{d^2 r_2}{a^2}
\left(e^{i\frac{2\pi}{6}\phi(\rv_1)}+e^{-i\frac{2\pi}{6}\phi(\rv_1)}\right)
\left(e^{i\frac{2\pi}{6}\phi(\rv_2)}+e^{-i\frac{2\pi}{6}\phi(\rv_2)}\right)
+\ldots\right] \\
&\prod_{n=1,2,3}\left[1 + e^{-E_{b}}\int \frac{d^2{r}_1}{a^2}
\left(e^{i\zh\times\grad\phi(\rv_1)\cdot\bv_n}+e^{-i\zh\times\grad\phi(\rv_1)\cdot\bv_n}\right)
\right.\\
& \left.+ \frac{1}{2!}e^{-2E_{b} }\int\frac{d^2 r_1}{a^2} 
\frac{d^2 r_2}{a^2}\left(e^{i\zh\times\grad\phi(\rv_1)\cdot\bv_n}
+e^{-i\zh\times\grad\phi(\rv_1)\cdot\bv_n}\right)
\left(e^{i\zh\times\grad\phi(\rv_2)\cdot\bv_n}
+e^{-i\zh\times\grad\phi(\rv_2)\cdot\bv_n}\right)
+\ldots\right],\\
\equiv &\int[d\phi]e^{-\tilde H},
\end{split}
\end{equation}
\end{widetext}
where non-neutral charge configurations vanish automatically after
integration over $\phi(\rv)$. In the last step, above, we have summed over only the positive/negative single charges of dislocation and disclination, and we obtain the dual vector sine-Gordon Hamiltonian,
\begin{equation}
\begin{aligned}
\label{DualSineGordon}
\tilde{H}=&\int_{\rv} \left[\frac{1}{2}\tilde{C}^{-1}_{ij,kl}
 \partial_i \partial_j \phi \partial_k\partial_l \phi\right.\\
&\left. -g_b \sum_{n=1}^{3} \cos({\bf b}_n\cdot\hat{\bf z}\times\grad\phi)
-g_s\cos\left(\frac{2\pi}{6} \phi\right)\right].
\end{aligned}
\end{equation}
Above, the couplings $g_b=\frac{2}{a^2} e^{-E_{b}}, g_s=\frac{2}{a^2}
e^{-E_{s}}$ are proportional to dislocation and disclination
fugacities, and the three elementary dislocation Burgers vectors are
given by ${\bf b}_1=a\hat{\bf x}, {\bf b}_2=-\frac{a}{2} \hat{\bf
  x}+\frac{a\sqrt{3}}{2}\hat{\bf y}, {\bf b}_3=-{\bf b}_1-{\bf
  b}_2=-\frac{a}{2} \hat{\bf x}-\frac{a\sqrt{3}}{2}\hat{\bf y}$.

\subsection{Vector sine-Gordon duality redux}

The above derivation of the dual vector sine-Gordon model departed from the conventional phonon-only elastic model of a 2D crystal,
Eq. \rf{H_el}. As discussed in Sec. \ref{sec:defects} target space
rotational invariance of the crystal is incorporated by building the
theory based on the {\em symmetric} tensor part $u_{ij}$,
Eq. \rf{harmonic_u} of the full strain tensor, $\partial_i u_j$, i.e., forbidding an explicit dependence on the local bond angle $\theta = \oh\epsilon_{ij}\partial_i u_j$, which corresponds to an angle of rotation of the crystal.

Alternatively, the rotational invariance of a crystal can be
formulated as a gauge-like (minimal) coupling between the full strain
tensor $\partial_i u_j$ and the bond angle $\theta({\bf r})$, encoded
in the Hamiltonian density
\begin{equation}
\begin{split}
\mathcal{H}&=\frac{1}{2} C_{ij,kl} (\partial_i u_j-\theta \epsilon_{ij}) (\partial_k u_l-\theta \epsilon_{kl})+\frac{1}{2} K (\partial_i \theta)^2.
\end{split}
\end{equation}
It can be straightforwardly verified that an integration over the
bond-angle field $\theta$ Higgs'es
out\cite{RadzihovskyLubensky,ChaikinLubensky} the antisymmetric
component of the strain tensor, at long wavelengths recovering the
conventional elastic Hamiltonian in \rf{H_el}.

We now decouple the strain and bond elastic terms by introducing two
Hubbard-Stratanovich fields -- the stress field $\sigma_{ij}$ and the
torque ``current'' $j_i$,
\begin{equation}
\begin{aligned}
    \mathcal{H}[{\bf u}, \theta; \sigma_{ij}, j_i]
=&\frac{1}{2} C^{-1}_{ij,kl} \sigma_{ij} \sigma_{kl}
+ i\sigma_{ij} (\partial_i u_j -\theta \epsilon_{ij}) \\
&+\frac{1}{2} K^{-1} j_i^2 +i j_i \partial_i \theta.
\label{Hgauge}
\end{aligned}
\end{equation}
We note that because $\partial_i u_j$ is not symmetrized, the stress
tensor $\sigma_{ij}$ is not symmetric here. In the presence of
topological defects, we again decompose the distortion field ${\bf u}$
and the bond angle $\theta$ into the smooth elastic and
nonsingle-valued components,
\begin{equation}
    u_i=\tilde{u}_i+u_i^s,\ \ \theta= \tilde{\theta}+\theta^s,
\end{equation}
which allow for dislocation and disclination defects, respectively.

Integrating out the single-valued parts $\tilde{{\bf u}}$ and
$\tilde{\theta}$ enforces two constraints
\begin{subequations}
\begin{eqnarray}
 \partial_i\sigma_{ij}=0,\\
 \partial_i j_i+\epsilon_{ij}\sigma_{ij}=0.
\end{eqnarray}
\end{subequations}
The first one is solved via a vector gauge field ${\bf A}$ with
\begin{equation}
  \sigma_{ij}=
  \epsilon_{ik}\partial_k A_j,
\end{equation}
which transforms the second constraint into
\begin{equation}
  \partial_i(j_i+A_i)=0.
\end{equation}
It is then solved by introducing another scalar potential $\phi$, via
$j_i=\epsilon_{ik} \partial_k \phi-A_i$. Expressing the Hamiltonian
\rf{Hgauge} in terms of gauge potentials, ${\bf A}({\bf r})$ and
$\phi({\bf r})$, integrating by parts and using the definitions of
dislocation ${\bf b}({\bf r})$ and disclination $s({\bf r})$
densities, the Hamiltonian density takes the form
\begin{equation}
\begin{split}
  \mathcal{H}=& \frac{1}{2} C^{-1}_{ij,kl} \epsilon_{im}
  \epsilon_{kn}\partial_m A_j \partial_n A_l+\frac{1}{2} K^{-1}
  (\epsilon_{ik} \partial_k\phi-A_i)^2\\
  &+i A_i b_i + i\phi\frac{2\pi}{p} s.
\end{split}
\label{Hutheta38}
\end{equation}
This model is evidently gauge-covariant under a local transformation,
\begin{subequations}
\begin{eqnarray}
{\bf A} ({\bf r}) &\to& {\bf A}({\bf r}) + \hat{{\bf z}}
\times \grad\alpha ({\bf r}),\\
\phi ({\bf r}) &\to& \phi ({\bf r})+
\alpha ({\bf r}). 
\end{eqnarray}
\end{subequations}
Integrating over the vector potential, ${\bf A}({\bf r})$ in the
partition function to lowest order yields
\begin{equation}
    A_i =\epsilon_{ik} \partial_k \phi,
\end{equation}
(an effective Higgs mechanism\cite{RadzihovskyLubensky,ChaikinLubensky}) and allows
us to eliminate ${\bf A}({\bf r})$ in favor of $\phi({\bf r})$ and to give the effective Hamiltonian density
\begin{equation}
\begin{split}
  \mathcal{H}[\phi]&=\frac{1}{2}
  \tilde{C}_{ij,kl}^{-1} \partial_i\partial_j
  \phi \partial_k\partial_l \phi
+ i \phi\left(\frac{2\pi}{p} s+\hat{\bf z}
  \cdot \grad \times {\bf b}\right),
\end{split}
\label{dualDefects1}
\end{equation}
which is identical to that found in \rf{Hphi}, which when supplemented by
dislocation and disclination core energies and summed over the defects
gives the dual vector sine-Gordon model, Eq. \rf{DualSineGordon}.

\subsection{Energetics of defects}

Inside the crystal state the background defects density vanishes. In
terms of the dual defects model, \rf{Hphi} and \rf{dualDefects1},
we can simply set the defect charges to zero, $s=\bv=0$. In terms of
the generalized sine-Gordon model, \rf{DualSineGordon}, this corresponds to the irrelevance of both cosines, i.e., vanishing couplings, $g_s=g_b=0$,
\begin{subequations}
\begin{eqnarray}
  H_{\text{cr}}&=&\frac{1}{2} \int_{\rv}
  \tilde{C}_{ij,kl}^{-1} \partial_i\partial_j
  \phi \partial_k\partial_l \phi,\\&=& \frac{1}{2} K^{-1} \int_{\rv} \left(\nabla^2 \phi \right)^2.
  \label{Hcrystal}
\end{eqnarray}
\end{subequations}
where $\tilde{C}^{-1}_{ij,kl}=\epsilon_{ia} \epsilon_{jb}
\epsilon_{kc} \epsilon_{ld} C^{-1}_{ab,cd}$, and we have specialized
it to that of a hexagonal lattice, obtaining \rf{Hcrystal} with
$K^{-1}=\frac{2\mu+\lambda}{4\mu(\mu+\lambda)}$.

The energy of a single disclination can be obtained by taking
$s(\rv)=2\pi\delta^2(\rv)$. Solving the corresponding Euler-Lagrange
equation for $\phi$ gives
\begin{equation}
\phi^{\text{cr}}_s(\kv) = \frac{i 2\pi K}{k^4}, 
\end{equation}
which for the energy of a single disclination in a crystal state gives
a well-known result,
\begin{subequations}
\begin{eqnarray}
E^{\text{cr}}_s &=& \oh K^{-1}\int d^2r (\nabla^2\phi_s)^2,\\
&=& \oh K \int_{L^{-1}} d^2 k\frac{1}{k^4}\sim K L^2,
\end{eqnarray}
\end{subequations}
where $L$ is the linear extent of the crystal. 

Similarly, for a single dislocation, such as $\bv (\rv)=\bv_1 \delta^2(\rv)=a\hat{\xv} \delta^2(\rv)$, the corresponding Euler-Lagrange equation for $\phi$ gives
\begin{equation}
    \phi_b^{\text{cr}}(\kv)= \frac{aK k_y}{k^4}=\frac{aK \sin \theta}{k^3},
\end{equation}
where $\theta$ is the angle between the direction of $\kv$ and the $\hat{\xv}$ axis. Therefore, the energy of a single dislocation in the crystal state is
\begin{subequations}
\begin{eqnarray}
     E_b^{\text{cr}}&=&\frac{1}{2} K^{-1} \int d^2 r (\nabla^2 \phi_c^{\text{cr}})^2,\\&=&\frac{1}{2} K a^2 \int_{L^{-1}}^{a^{-1}} \frac{d^2k}{(2\pi)^2} \frac{\sin^2\theta}{k^2},\\&=&\frac{1}{8} Ka^2 \ln \frac{L}{a} \sim K\ln L.
\label{Dislocationenergy}
\end{eqnarray}
\end{subequations}
The $C_3$ rotational symmetry guarantees that the energies for the
other single dislocations, $\bv (\rv)=\bv_2 \delta^2(\rv)$ and $\bv
(\rv)=\bv_3 \delta^2(\rv)$ are identical.

\section{Renormalization-group analysis of the melting transition}
\label{sec:RGsineGordon}

As discussed in the Introduction, and calculated above, within the
crystal state with a vanishing background defect density, the energy
of a single dislocation scales as $E_{b}\sim\ln L$, while the energy
of a single disclination scales as $E_{s}\sim L^2$, where $L$ is the
linear extent of the crystal. Thus, as discovered by Nelson and
Halperin\cite{Nelson 79}, above the critical melting temperature
$T_m$, the dislocations will unbind first, while the disclinations
remain confined, leading to the orientationally ordered hexatic
liquid, which is stable in a finite temperature range $T_m < T < T_{\text{hex}}$.

More formally, within the dual sine-Gordon model this is reflected by
the irrelevance of the disclination cosine operator at the Gaussian
fixed line. Computing its average in a system of size $L$, we indeed
find
\begin{subequations}
\begin{eqnarray}
\left\langle\int d^2r
\cos\left(\frac{2\pi}{6}\phi\right)\right\rangle&=&\int d^2 r  e^{-\frac{\pi^2}{18} \langle \phi^2(\rv) \rangle},\\&=& \int d^2r  e^{-\frac{\pi }{72}K L^2} ,\\
&\sim& L^2 e^{-L^2}\rightarrow 0.
\end{eqnarray}
\end{subequations}
This analysis [that can be more formally elevated to a renormalizaiton
group (RG) computation] demonstrates that the disclination cosine
operator, $g_s$, is strongly irrelevant around the Gaussian fixed line,
i.e., when the dislocation cosine, $g_b$, is small, corresponding to the
absence of screening of disclinations by dislocations.

\subsection{Crystal-hexatic melting transition}
Thus, within the crystal and near the crystal-to-hexatic transition,
we can neglect the disclination cosine, setting $g_s = 0$, reducing
the effective Hamiltonian to
\begin{equation}
\begin{aligned}
    \tilde{H}_{\text{cr}}=\int_{\rv}\left[\frac{1}{2}
      \tilde{C}^{-1}_{ij,kl} \partial_i \partial_j
      \phi \partial_k\partial_l \phi -g_b \sum_{n=1}^{3} \cos({\bf
        b}_n \cdot \hat{\bf z}\times\grad\phi)\right].\;\;\;\;\;
\end{aligned}
\end{equation}

The RG analysis of this model is more convenient in an equivalent
description in terms of a divergenceless vector field ${\bf A}=\hat{\bf
  z}\times\grad\phi$,
\begin{equation}
\begin{aligned}
  \tilde{H}_{\text{cr}}=&\int_{\rv} \left[\frac{1}{2} C^{-1}_{ij,kl}
  \epsilon_{im}\epsilon_{kn} \partial_m A_j \partial_n A_l\right.\\
  &\left.+\frac{\alpha}{2} \left(\grad\cdot {\bf A}\right)^2 -g_b \sum_{n=1}^{3}
  \cos({\bf b}_n \cdot {\bf A})\right],
\end{aligned}
\end{equation}
with the constraint $\grad\cdot{\bf A}=0$ imposed energetically via a
``mass'' term $\frac{\alpha}{2} \left(\grad\cdot{\bf A}\right)^2$ added to the
Hamiltonian, with $\alpha \to \infty$ taken at the end of the
calculation.  Interestingly, our model is mathematically closely
related to that for the roughenning transition of a crystal pinned by
a commensurate substrate, studied by Ohta\cite{Ohta80}, and by Levin and Dawson\cite{Levin 90}.

Specializing $C^{-1}_{ij,kl}$ to a hexagonal lattice, Eq. \rf{Cinv}, the
Hamiltonian reduces to
\begin{equation}
\begin{aligned}
  \tilde{H}_{cr}=&\int_{\rv} \left[\frac{K^{-1}}{2}
 \left(\partial_i A_j\right)^2
+ \frac{B}{2} \partial_i A_j \partial_j A_i+ \frac{\alpha}{2} \left(\grad\cdot {\bf A}\right)^2\right.\\&
\left.-g_b \sum_{n=1}^{3}\cos({\bf b}_n \cdot {\bf A})\right],
\end{aligned}
\end{equation}
where the couplings are
\begin{subequations}
\begin{eqnarray}
K^{-1}&=&\frac{2\mu+\lambda}{4\mu(\mu+\lambda)},\\
B &=&\frac{\lambda}{4\mu(\mu+\lambda)}.
\end{eqnarray}
\end{subequations} 
In the physical limit $\alpha \to \infty$, the dislocation-free, Gaussian
propagator is given by,
\begin{equation}
  \langle A_i({\bf q}) A_j ({\bf q}')\rangle_0
  =\frac{K}{q^2}P^T_{ij}(\qv)(2\pi)^2\delta^2(\qv+\qv'),
\end{equation}
a purely transverse form, with the transverse projection operator,
$P_{ij}^T(\qv) = \delta_{ij}- \frac{q_iq_j}{q^2}$ consistent
with \rf{Kij}, encoding the target-space rotational invariance of the
crystal\cite{Nelson 79,Young 79}.

To describe the melting transition we need to include dislocations,
encoded in the $g_b$ cosine operator. Although at low temperature,
(corresponding to large elastic constants and small $K^{-1}$) a
perturbative expansion in $g_b$ is convergent, (i.e., the fixed line
$g_b = g_s = 0$ is stable), it breaks down for $K$ below a critical
value, indicating an entropic proliferation of large dislocation pairs
for $T > T_m$.

To treat this high-temperature nonperturbative regime requires an RG
analysis. Relegating the details to Appendix A, here we present the
highlights of the analysis and its results. To control the divergent
perturbation theory, we employ the momentum-shell coarse-graining RG
by decomposing the vector field ${\bf A}({\bf r})$ into its
short-scale and long-scale modes, $A_i({\bf r})=A_i^{<}({\bf
  r})+A_i^{>}({\bf r})$, with
\begin{subequations}
\begin{eqnarray}
  A_i^{<}({\bf r})=\int _{0<q<\Lambda/b} \frac{d^2q}{(2\pi)^2} e^{i
    {\bf q}\cdot {\bf r}} A_i({\bf q}),\\
  A_i^{>}({\bf r})=\int _{\Lambda/b<q<\Lambda} \frac{d^2q}{(2\pi)^2}
  e^{i {\bf q}\cdot {\bf r}} A_i({\bf q}),
\end{eqnarray}
\end{subequations}
where the ultra-violet cutoff $\Lambda=2\pi/a$, and the rescaling
factor $b > 1$ defines the width of the momentum shell, $\Lambda/b < q
< \Lambda$.  Following a standard analysis, we integrate short scale
modes $A_i^{>}({\bf r})$ out of the partition function, obtaining a
coarse-grained Hamiltonian for the long-scale modes, $A_i^{<}({\bf
  r})$, with the renormalized coupling $K_R^{-1}(b)$, $B_{R}(b)$ and
$g_{b_R}(b)$ satisfying
\begin{subequations}
\begin{eqnarray}
  K_R^{-1}(b)&=& K^{-1} + J_2 g_b^2 ,\\ 
  B_{_R}(b)&=&B+J_3 g_b^2,\\
  g_{b_R} (b)&=&g_b e^{-\frac{1}{2} G_{nn}^{>}(0)} + J_1 g_b^2,\label{gb}
\end{eqnarray}
\end{subequations}
valid to second-order in $g_b$. The Greens function appearing 
above is given by
\begin{equation}
G^>_{nm}({\bf r}_1-{\bf r}_2)\equiv b_i^n b_j^m \langle A_i^{>}({\bf r}_1)
A_j^{>} ({\bf r}_2) \rangle_0^{>},
\end{equation}
and $J_i$ factors are defined as,
\begin{subequations}
\begin{eqnarray}
J_1&=&\pi a^2\left[e^{\frac{\overline{K}}{16\pi}}I_0\left(\frac{\overline{K}}{8\pi}\right)
+\left(\frac{\overline{K}}{16\pi}-1\right)\right]\ln b,\\
J_2&=&\frac{\pi a^6}{4}
\left[e^{\frac{\overline{K}}{8\pi}}\left(\frac{3}{2}I_0\left(\frac{\overline{K}}{8\pi}\right)
-\frac{3}{4}I_1\left(\frac{\overline{K}}{8\pi}\right)\right)\right.\nonumber\\
&&\left.+\frac{3}{2}\left(\frac{\overline{K}}{16\pi}-1\right)\right]\ln b,\\
J_3&=&\frac{3\pi a^6}{16}e^{\frac{\overline{K}}{8\pi}}I_1\left(\frac{\overline{K}}{8\pi}\right)\ln b,
\end{eqnarray}
\end{subequations}
where $I_0(x)$ and $I_1(x)$ are modified Bessel functions.

It is convenient to examine an infinitesimal form of these RG
equations by taking $b=e^{\delta l}$ with $\delta l\ll 1$.  Near the
melting critical point $K_R^{-1}(l \to \infty)\equiv
K_{R*}^{-1}=\frac{a^2}{16\pi}, g_{b_R}(l \to \infty)\equiv g_{b}^*=0$,
this then gives RG differential flow equations for the dimensionless
coupling constants $\overline{K}^{-1}(l)=K^{-1}/a^2$, $\overline{B}(l)=B/a^2$ and $\overline{g}_b(l)=g_b a^2$ ,
\begin{subequations}
\begin{eqnarray}
  \frac{d\overline {K}^{-1}(l)}{dl}&=&\frac{3\pi}{8} \left[e^2 \left(I_0(2)-\frac{1}{2} I_1(2)\right) \right] \cdot \overline{g}_b^2(l) ,\;\;\;\;\;\;\\
  \frac{d\overline{B}(l)}{dl}&=&\frac{3\pi}{16} e^2 I_1(2) \cdot \overline{g}_b^2(l),\\
  \frac{d \overline{g}_b(l)} {dl}&=& \left(2-\frac{\overline{K}}{8\pi}\right) \overline{g}_b+ \pi e I_0(2) \cdot \overline{g}_b^2(l).  
\end{eqnarray}
\end{subequations}
Using the definitions in terms of the dimensionless Lam\'e elastic
constants $\overline{\mu}=\mu a^2$, $\overline{\lambda}=\lambda a^2$,
and the fugacity $y$, 
\begin{subequations}
\begin{eqnarray}
\overline{K}^{-1}&=&\frac{1}{4}\left(\frac{1}{\overline{\mu}}+\frac{1}{\overline{\mu}+\overline{\lambda}} \right),\\
\overline{B}&=&\frac{1}{4} \left(\frac{1}{\overline{\mu}}-\frac{1}{\overline{\mu}+\overline{\lambda}} \right),\\
\bar{g}_b&=&2 e^{-E_{b}}=2 y,  
\end{eqnarray}
\end{subequations}
our equations reduce exactly to the seminal RG flows for the inverse
shear modulus, $\overline{\mu}^{-1}(l)$, the inverse bulk modulus $[\overline{\mu} (l)
+\overline{\lambda}(l)]^{-1}$, and the effective fugacity $y(l)$, respectively,
\begin{subequations}
\begin{eqnarray}
  \frac{d\overline{\mu}^{-1}}{dl}&=& 3\pi e^2 I_0(2) y^2 ,\\
  \frac{d (\overline{\mu}+\overline{\lambda})^{-1}} {dl}&=& 3\pi e^2 \left[I_0(2)-I_1(2)\right] y^2,\\
  \frac{d y} {dl}&=& \left(2-\frac{\overline{K}}{8\pi}\right)y +2\pi e I_0(2)y^2,
\end{eqnarray}
\end{subequations}
first derived by Nelson and Halperin\cite{Nelson 79}, and
Young\cite{Young 79}.

Following a standard analysis\cite{Kosterlitz 72,Nelson 79}, the
characteristic correlation length $\xi_{\text{xtal-hex}}$ near the critical
point at $T\to T_m^-$ can be extracted from the above RG flows, and it is given by
\begin{equation}
\xi_{\text{xtal-hex}}(T) \sim a e^{-c/|T-T_m|^{\overline{\nu}}},
\end{equation}
with the hexagonal lattice exponent given by
\begin{equation}
\overline{\nu}=0.3696\ldots,
\label{nu}
\end{equation}
and $c$ a nonuniversal constant\cite{Nelson 79}.

\subsection{Hexatic-isotropic liquid transition}

Dislocation-unbinding above the melting temperature, $T_m$ destroys
the crystal order, restoring continuous translational symmetry. The
plasma of unbound dislocations drives the shear modulus to zero, but
retains the quasi-long ranged orientational order and the associated
bond orientatonal stiffness. Inside this orientationally-ordered
hexatic fluid ($T_m < T < T_{hex}$) $g_b$ is driven to strong coupling,
suppressing ${\bf A}$ fluctuations, and allowing us to approximate
the dislocation cosine by its harmonic form. With disclinations
reinstated, the resulting effective Hamiltonianin takes the standard
scalar sine-Gordon form:
\begin{equation}
\begin{split}
   \tilde{H}_{\text{hex}} &\approx \int_{\rv} \left[\frac{1}{2} g_b
     \sum_{n=1}^3 \epsilon_{ik} \epsilon_{jl} b^n_i b^n_j\partial_k
     \phi \partial_l \phi-g_s \cos\left(\frac{2\pi}{6} \phi \right)
   \right]\\
   &=\int_{\rv} \left[\oh J (\nabla \phi)^2 - 
g_s\cos\left(\frac{2\pi}{6} \phi \right) \right].
\end{split}
\end{equation}
where $J\equiv\frac{3}{2} a^2 g_b$. 

Alternatively, we can get to this dual hexatic Hamiltonian by noting
that above the critical melting temperature, $T_m$, dislocations
(dislination dipoles) unbind, leading to an orientationally ordered (a
hexatic) fluid. Since the dislocations then appear at finite density,
their Burgers charge, $\bv(\rv)$ can be treated as a continuous
(rather than a discrete) vector field. Going back to \rf{Hdiscrete},
carrying out a Gaussian integral over a continuous field $\bv_\rv$, and
summing over discrete disclination charges, $s_{\rv_n}$ we again
obtain the hexatic Hamiltonian presented above.
  
Utilizing $\tilde{H}_{hex}$ we observe that within the hexatic phase,
the energy of a disclination, screened by the plasma of proliferated
dislocations, is reduced significantly from that of the crystal (where
it diverges as $L^2$) to $E_s^{hex} \sim J_R\ln L/a$. Thus, the
hexatic-isotropic fluids transition is of the conventional
Kosterlitz-Thouless type\cite{Kosterlitz 72,Nelson 79,ChaikinLubensky}, taking place
at $T_{hex} =(72/\pi)J_R(T_{hex})$. Above $T_{hex}$ the fluid is
isotropic, characterized by short-ranged translational and
orientational correlations.

\section{Summary and conclusion}
\label{sec:summary}
In this paper, starting with a descripton of a crystal in terms of its
elasticity and topological defects, we derive a corresponding dual
vector sine-Gordon model. In the latter, the disclinations and
dislocations are captured by cosine operators of the Airy stress
potential and its gradient. The relevance of the latter dislocation
cosine signals the continuous
Kosterlitz-Thouless-Halperin-Nelson-Young melting transition of a
crystal into a hexatic fluid\cite{Kosterlitz 72, Nelson 79, Young
  79}. The subsequent relevance of the former disclination cosine
captures the hexatic-to-isotropic fluid Kosterlitz-Thouless
transition, as outlined in Sec. \ref{sec:sketch} and illustrated in
Fig. \ref{fig:RGflows}. Our complementary analysis reproduces straightforwardly the results of Nelson and Halperin\cite{Nelson 79} and Young\cite{Young 79}, including the correlation functions, defects
energetics, renormalization-group flows, and the correlation length
exponent $\overline{\nu}$.

We expect that the simplified vector sine-Gordon formulation,
presented here will be useful in further detailed studies of, e.g.,
the external stress, the defects dynamics, the substrate, and the dynamics of the melting transition\cite{dynamicsMeltingPRB80}.

\acknowledgments

We acknowledge useful discussions with Michael Pretko, and we thank him
for valuable input on the manuscript. This work was supported by the
Simons Investigator Award from the James Simons Foundation and by NSF MRSEC Grant No. DMR-1420736.  L.R. also thanks the KITP for its
hospitality as part of the Fall 2016 Synthetic Matter workshop and
sabbatical program, during which this work was initiated and supported
by NSF Grant No. PHY-1125915.

\appendix
\section{ Derivation of RG equations}
\label{appendixA}
In this appendix we present the details of the renormalization-group
analysis of the vector sine-Gordon model for the dislocation unbinding
transition,
\begin{equation}
\begin{aligned}
  \tilde{H}=&\int_{\rv} \left[\right.\frac{1}{2} \left(K^{-1}
    \left(\partial_i A_j\right)^2+B \partial_i A_j \partial_j A_i
     \right)+\frac{\alpha}{2}\left(\grad \cdot {\bf A}\right)^2\\& -g_b \sum_{n=1}^{3}
  \cos({\bf b}_n \cdot {\bf A})\left. \right],
\end{aligned}
\end{equation}
where the coupling constants are $K^{-1}=
\frac{2\mu+\lambda}{4\mu(\mu+\lambda)}$ and $B=\frac{\lambda}{4\mu(\mu+\lambda)}$. The transversality
constraint $\nabla \cdot {\bf A}=0$ is imposed energetically by taking
$\alpha \to \infty$ at the end of the calculation.

In the physical limit $\alpha \to \infty$, the dislocation-free,
Gaussian propagator is given by,
\begin{equation}
  \langle A_i({\bf q}) A_j ({\bf q}') \rangle_0=  \frac{K}{q^2}P^T_{ij}(\qv)(2\pi)^2\delta^2(\qv+\qv'),
\end{equation}
with the transverse projection operator, $P_{ij}^T(\qv) = \delta_{ij}-
\frac{q_iq_j}{q^2}$.

To carry out the momentum-shell RG analysis, we decompose vector field
${\bf A}({\bf r})$ into its short- and long-scale modes, $A_i({\bf
  r})=A_i^{<}({\bf r})+A_i^{>}({\bf r})$, with
\begin{subequations}
\begin{eqnarray}
  A_i^{<}({\bf r})=\int _{0<q<\Lambda/b} \frac{d^2q}{(2\pi)^2} e^{i
    {\bf q}\cdot {\bf r}} A_i({\bf q}),\\
  A_i^{>}({\bf r})=\int _{\Lambda/b<q<\Lambda} \frac{d^2q}{(2\pi)^2}
  e^{i {\bf q}\cdot {\bf r}} A_i({\bf q}),
\end{eqnarray}
\end{subequations}
where the ultra-violet cutoff $\Lambda=2\pi/a$, and the rescaling
factor $b > 1$ defines the width of the momentum shell, $\Lambda/b < q
< \Lambda$.

Integrating out the short-scale modes, $A_i^{>}({\bf r})$, the
partition function reduces to integration over the long-scale
modes, $A^{<}_i({\bf r})$ with an effective Hamiltonian,
\begin{equation}
  \begin{split}
  Z&=\int [d{\bf A}] e^{-H[{\bf A}^{>}+{\bf A}^{<}]}\\
&=\int [d{\bf A}^{<}] [d{\bf A}^{>}]
e^{-H_0[{\bf A}^{<}]-H_0[{\bf A}^{>}]-H_{\text{int}}[{\bf A}^{<}+{\bf A}^{>}]} \\
&= \int[d{\bf A}^{<}]e^{-H_0[{\bf A}^{<}]} Z_0^{>} \langle
e^{-H_{\text{int}}[{\bf A}^{<}+{\bf A}^{>}]} \rangle_0^{>}\\
&\equiv \int [d{\bf A}^{<}]e^{-H_<[{\bf A}^{<}]},
\end{split}
\end{equation}
where $Z_0^{>}=\int [d{\bf A}^>] e^{-H_0[{\bf A}^>]}$ is the harmonic
part of the partition function of the short-scale modes with the
quadratic Hamiltonian,
\begin{equation}
\begin{split}
      H_0[{\bf A}^{>}]=& \int_{\rv} \left[\right.\frac{1}{2}\left( K^{-1}
    \left(\partial_i A^{>}_j\right)^2+B \partial_i A^>_j \partial_j A^>_i \right)\\ &+\frac{\alpha}{2}
    \left(\grad \cdot {\bf A}^{>}\right)^2 \left. \right],
\end{split}
\end{equation}
and, the coarse-grained effective Hamiltonian, $H_<[{\bf A}^{<}]$ of
the long-scale modes given by,
\begin{eqnarray}
    H_<[{\bf A}^{<}]=H_0[{\bf A}^{<}]-\ln \langle e^{-H_{\text{int}}[{\bf A}^{<}+{\bf A}^{>}]} \rangle_0^{>}-\ln Z_0^{>}.
\end{eqnarray}
We drop the last term, $-\ln Z_0^>$, which is a field independent
correction to the free energy that does not affect the flow of the coupling
constants.  We then compute $H_<[{\bf A}^{<}]$ in terms of corrections to
elastic constants $\mu,\lambda$ and dislocation fugacity $g_b$,
arising from coarse-graining $-\ln\langle
e^{-H_{\text{int}}[{\bf A}^{<}+{\bf A}^{>}]} \rangle_0^{>}$.

We expand $\langle e^{-H_{\text{int}}[{\bf A}^{<}+{\bf A}^{>}]}
\rangle_0^{>}$ to second order in $g_b$,
\begin{equation}
  \begin{split}
   &\langle e^{-H_{\text{int}}[{\bf A}^{<}+{\bf A}^{>}]} \rangle_0^{>} = \langle e^{
     g_b \sum_{n=1,2,3} \int_{\rv} \cos ({\bv}_n \cdot {\bf A})} \rangle_0^{>}\\
&\approx  1+g_b \sum_{n=1}^3\int_{\rv} \langle\cos ({\bv}_n \cdot {\bf A}) \rangle_0^{>}\\&+ \frac{g_b^2}{2} \sum_{n=1}^3 \sum_{m=1}^3 \int_{\rv_1} \int_{\rv_2}\langle\cos ({\bv}_n \cdot {\bf A}({\bf r}_1)) \cos ({\bv}_m \cdot {\bf A}({\bf r}_2)) \rangle_0^{>},
  \end{split}
\end{equation}
finding,
\begin{equation}
\begin{aligned}
  &\ln\langle e^{-H_{\text{int}}[{\bf A}^{<}+{\bf A}^{>}]} \rangle_0^{>}=g_b \sum_{n=1}^3\int_{\rv} \langle \cos ({\bv}_n \cdot {\bf A}) \rangle_0^{>}\\&+ \frac{g_b^2}{2} \sum_{n,m=1}^3 \int_{\rv_1} \int_{\rv_2}\left[\langle \cos ({\bv}_n \cdot {\bf A} ({\bf r}_1)) \cos ({\bv}_m \cdot {\bf A}({\bf r}_2)) \rangle_0^{>} \right.\\&\left.-\langle \cos ({\bv}_n \cdot {\bf A}({\bf r}_1)) \rangle_0^{>} \langle \cos ({\bv}_m \cdot {\bf A}({\bf r}_2)) \rangle_0^{>} \right].
\end{aligned}
\end{equation}
These are straightforwardly evaluated by Gaussian integration, giving,
to first order
 \begin{eqnarray}
  \begin{aligned}
   \langle \cos ({\bv}_n \cdot {\bf A}) \rangle_0^{>} &=  \frac{1}{2} \langle e^{i{\bf b}_n \cdot ({\bf A}^{<}+{\bf A}^{>})}+ e^{-i{\bf b}_n \cdot ({\bf A}^{<}+{\bf A}^{>})} \rangle_0^{>}\\& = e^{-\frac{1}{2} \langle ({\bf b}_n \cdot {\bf A}^{>})^2 \rangle_0^{>}} \cos \left({\bv}_n \cdot {\bf A}^{<}\right)\\&\equiv  e^{-\frac{1}{2} G^{>}_{nn}(0)} \cos \left({\bv}_n \cdot {\bf A}^{<}\right),
 \end{aligned}
 \end{eqnarray}
 and to second order the connected part,
 \begin{widetext}
 \begin{eqnarray}
 \begin{aligned}
   &\langle\cos( {\bv}_n \cdot {\bf A}({\bf r}_1)) \cos({\bv}_m \cdot
   {\bf A} ({\bf r}_2)) \rangle_0^{>}-\langle \cos({\bv}_n \cdot {\bf
     A}({\bf r}_1))\rangle_0^{>} \langle \cos ({\bv}_m \cdot {\bf
     A}({\bf r}_2))\rangle_0^{>}\\=& \frac{1}{4} \left \{
     e^{i\left({\bv}_n \cdot {\bf A}^{<}({\bf r}_1) +{\bv}_m \cdot
         {\bf A}^{<} ({\bf r}_2) \right)} \langle e^{i\left({\bv}_n
         \cdot {\bf A}^{>}({\bf r}_1) +{\bv}_m \cdot {\bf A}^{>} ({\bf
           r}_2) \right)} \rangle_0^{>} +e^{i\left({\bv}_n \cdot {\bf
           A}^{<}({\bf r}_1) -{\bv}_m \cdot {\bf A}^{<} ({\bf r}_2)
       \right)} \langle e^{i\left({\bv}_n \cdot {\bf A} ^{>}({\bf
           r}_1) -{\bv}_m \cdot {\bf A}^{>} ({\bf r}_2) \right)}
     \rangle_0^{>} \right.\\+&\left.  e^{-i\left({\bv}_n \cdot {\bf
           A}^{<}({\bf r}_1) -{\bv}_m \cdot {\bf A}^{<} ({\bf r}_2)
       \right)} \langle e^{-i\left({\bv}_n \cdot {\bf A}^{>}({\bf
           r}_1) -{\bv}_m \cdot {\bf A}^{>} ({\bf r}_2) \right)}
     \rangle_0^{>}+ e^{-i\left({\bv}_n \cdot {\bf A}^{<}({\bf r}_1)
         +{\bv}_m \cdot {\bf A}^{<} ({\bf r}_2) \right)} \langle
     e^{-i\left({\bv}_n \cdot {\bf A} ^{>}({\bf r}_1) +{\bv}_n \cdot
         {\bf A}^{>} ({\bf r}_2) \right)} \rangle_0^{>} \right\}\\-&
   e^{-G_{nn}^{>}(0)} \cos ({\bv}_n \cdot {\bf A}^{<}({\bf r}_1)) \cos
   ({\bv}_m \cdot {\bf A}^{<} ({\bf r}_2))\\
=&\oh e^{- G_{nn}^{>}(0)}\left \{ \left[ e^{- G_{nm}^{>}({\bf v})}-1\right] \cos  \left[ {\bv}_n \cdot {\bf A}^{<}({\bf r}_1) +{\bv}_m \cdot {\bf A}^{<} ({\bf r}_2) \right]+\left[ e^{+ G_{nm}^{>}({\bf v})}-1\right] \cos  \left[{\bv}_n \cdot {\bf A}^{<}({\bf r}_1) -{\bv}_m \cdot {\bf A}^{<} ({\bf r}_2) \right] \right\},
\end{aligned}
\end{eqnarray}
\end{widetext}
where ${\bf v}={\bf r}_1-{\bf r}_2$, and we have defined
 \begin{equation}
  G_{nm}^>({\bf r}_1-{\bf r}_2)\equiv b_i^n b_j^m \langle A_i^{>}({\bf r}_1) A_j^{>} ({\bf r}_2) \rangle_0^{>}.
 \end{equation}
 For $0<v<a$, we approximate the short-scale averaged correlation function by its value at $v=0$,
 \begin{equation}
     \begin{split}
     \langle A_i^{>}({\bf r}) A_j^{>} ({\bf r}) \rangle_0^{>} &=\int_{\bf q}^{>} \langle A_i({\bf q}) A_j(-{\bf q})\rangle_0^{>}\\&=  \int_{\Lambda/b}^{\Lambda} \frac{q dq }{(2\pi)^2}\int_0^{2\pi} d\theta    \frac{K}{q^2} \left( \delta_{ij}-\frac{q_i q_j}{q^2} \right) \\&=\frac{K}{2\pi} \ln b \cdot \left(\delta_{ij}-\frac{1}{2} \delta_{ij} \right)\\&= \frac{K}{4\pi}\ln b \cdot  \delta_{ij}.
     \end{split}
  \end{equation}
 For $v>a$, we have the real space correlation function\cite{Nelson 79},
  \\
  \\
  \\
  \\
\begin{equation}
\begin{split}
   \langle A_i({\bf r}_1) A_j ({\bf r}_2) \rangle &=\int \frac{d^2
     q}{(2\pi)^2} \langle A_i({\bf q}) A_j (-{\bf q}) \rangle \cdot
   e^{i{\bf q}\cdot({\bf r}_1-{\bf r}_2)}\\
&=\int \frac{d^2 q}{(2\pi)^2} \frac{K}{q^2}
\left(\delta_{ij}-\frac{q_i q_j}{q^2} \right) \cdot e^{i{\bf q} \cdot
  {\bf v}}\\
&=-\frac{K}{4\pi} \left(\ln \frac{v}{a} \cdot \delta_{ij}-\frac{v_i v_j}{v^2}\right).
\end{split}
\end{equation}
Therefore, we can evaluate $G^>_{nm}({\bf r}_1-{\bf r}_2)$ explicitly, finding
\begin{equation}
\begin{split}
&G^{>}_{nm}({\bf r}_1-{\bf r}_2) \approx\\
& \left\{\begin{array}{ll}
\frac{K}{4\pi} {\bf b}_n \cdot {\bf b}_m \ln b, \mbox{ for $ 0 < v <a$},
\\-\frac{K}{4\pi} \left[ {\bf b}_n \cdot {\bf b}_m \ln \frac{v}{ba}
  -\frac{({\bf b}_n \cdot {\bf v})({\bf b}_m \cdot {\bf
      v})}{v^2}\right], 
\mbox{for $ a < v < ba$}, \\
 0, \mbox{ for $v>ba$.} \end{array}\right. 
    \end{split}
\end{equation}
This thus gives to second-order in $g_b$,
\begin{widetext}
\begin{equation}
\begin{aligned}
  \ln\langle e^{-H_{\text{int}}[{\bf A}^{<}+{\bf A}^{>}]} \rangle_0^{>}=&g_b \sum_{n} e^{-\frac{1}{2}  G_{nn}^{>}(0)} \int_{\rv} \cos \left({\bv}_n \cdot {\bf A}^{<}\right) \\&+\frac{g_b^2}{4} \sum_{n,m}  e^{-G_{nn}^{>}(0)} \left \{\int_{\rv_1} \int_{\rv_2} \left[ e^{- G_{nm}^{>}({\bf r}_1-{\bf r}_2)}-1\right]\cos \left({\bv}_n \cdot {\bf A}^{<}({\bf r}_1) +{\bv}_m \cdot {\bf A}^{<} ({\bf r}_2) \right)\right.\\& \left.+ \int_{\rv_1} \int_{\rv_2} \left[ e^{+ G_{nm}^{>}({\bf r}_1-{\bf r}_2)}-1\right] \cos \left({\bv}_n \cdot {\bf A}^{<}({\bf r}_1) -{\bv}_m \cdot {\bf A}^{<} ({\bf r}_2) \right) \right \}.
  \label{lnH_int}
\end{aligned}
\end{equation}
\end{widetext}

The above double integral, $\int d^2 r_1 \int d^2 r_2 (...)$ simplifies
using the fact that $G_{nm}^{>}({\bf r}_1-{\bf r}_2)$ is short-ranged,
vanishing for $|{\bf r}_1-{\bf r}_2|$ larger than $b/\Lambda \sim ba$,
since $G_{nm}^{>}({\bf r})$ is defined to be composed of Fourier modes
only within a thin momentum-shell, $\Lambda/b < q < \Lambda$.
Consequently, $\left[ e^{\pm G_{nm}^{>}({\bf r}_1-{\bf
      r}_2)}-1\right]$ is also small everywhere but in the range,
$|{\bf r}_1-{\bf r}_2| \sim b/\Lambda \sim ba $. To utilize these
observations we change variables $\{{\bf r}_1, {\bf r}_2\}$ to a their
sum and difference,
\begin{equation}
 {\bf r}=\frac{1}{2} ({\bf r}_1+{\bf r}_2),\;\; {\bf v} ={\bf r}_1-{\bf r}_2,
\end{equation}
obtaining
\begin{equation}
   {\bv}_n \cdot {\bf A}^{<}({\bf r}_1) +{\bv}_m \cdot {\bf A}^{<} ({\bf r}_2) \approx ({\bf b}_n+{\bf b}_m)\cdot {\bf A}^{<}({\bf r}),
\end{equation}
and,
\begin{equation}
\begin{aligned}
 &{\bv}_n\cdot {\bf A}^{<}({\bf r}_1) -{\bv}_m \cdot {\bf A}^{<} ({\bf r}_2)=
{\bf b}_n\cdot{\bf A}^{<}({\bf r}+\frac{{\bf v}}{2})
-{\bf b}_m\cdot {\bf A}^{<}({\bf r}-\frac{{\bf v}}{2}) \\
&\approx ({\bf b}_n-{\bf b}_m)\cdot {\bf A}^{<}({\bf r})+\oh({\bf b}_n+{\bf b}_m)\cdot({\bf v} \cdot \nabla) {\bf A}^{<}({\bf r}).
\end{aligned}
\end{equation}
This allows the following simplifications of \rf{lnH_int},
\begin{widetext}
\begin{equation}
\begin{aligned}
  \ln\langle e^{-H_{\text{int}}[{\bf A}^{<}+{\bf A}^{>}]} \rangle_0^{>}
  \approx & g_b \sum_{n} e^{-\frac{1}{2} G_{nn}^{>}(0)} \int_{\rv}
  \cos
  \left({\bf b}_n\cdot {\bf A}^{<}\right)\\
  &+\frac{g_b^2}{4} \sum_{n,m} e^{- G_{nn}^{>}(0)}\left \{\int_{\rv}
    \int_{\bf v}\left[ e^{- G_{nm}^{>}({\bf v})}-1\right]\cos
    \left(({\bf b}_n+{\bf b}_m)\cdot{\bf A}^{<}({\bf r}) \right) \right.\\
  & \left.+\left[ e^{+ G_{nm}^{>}({\bf v})}-1\right] 
\cos\left[\left(({\bf b}_n-{\bf b}_m)+\oh({\bf b}_n+{\bf b}_m)
({\bf v}\cdot \nabla)\right)\cdot{\bf A}^{<} ({\bf r}) \right] \right
  \}.
  \label{lnH_int_new}
\end{aligned}
\end{equation}
\end{widetext}
Comparing \rf{lnH_int_new} to the component of the long-scale
Hamiltonian $H[{\bf A}^{<}]$, we can extract the corresponding corrections
for the coupling constants $K^{-1}, B$, and $g_b$. To this end,
ignoring the field-independent terms, we obtain
\begin{widetext}
 \begin{equation}
 \begin{aligned}
   \ln\langle e^{-H_{\text{int}}[{\bf A}^{<}+{\bf A}^{>}]}
   \rangle_0^{>}\approx & g_b \sum_{n} e^{-\frac{1}{2} G^{>}(0)} \int_{\rv} \cos \left({\bf b}_n\cdot {\bf A}^{<}\right) \\
   &+\frac{g_b^2}{4} \sum_{n,m} e^{- G^{>}(0)} \left \{\int_{\rv} \int_{\bf v}\left[ e^{- G_{nm}^{>}({\bf v})}-1\right] \cos \left(({\bf
         b}_n+{\bf b}_m)\cdot{\bf A}^{<}({\bf r}) \right)
   \right.\\
   &\left.+ \int_{\rv} \int_{\bf v} \left[ e^{+ G_{nm}^{>}({\bf
           v})}-1\right] 
\left[ \cos \left(({\bf b}_n-{\bf b}_m)\cdot{\bf A}^{<} \right) 
-\frac{1}{8}\left(({\bf b}_n+{\bf b}_m)\cdot({\bf v}\cdot \nabla)
{\bf A}^{<} \right)^2
\cos\left(({\bf b}_n-{\bf b}_m)\cdot{\bf A}^{<} \right) \right]\right\}\\
=& g_b \sum_{n} e^{-\frac{1}{2} G^{>}(0)} \int_{\rv} \cos
\left({\bf b}_n\cdot{\bf A}^{<}\right) \\
&+g_b^2 \int_{\rv} 
\left[I_{23}\cos\left({\bf b}_1\cdot{\bf A}^{<}({\bf r}) \right)
+I_{13}\cos \left({\bf b}_2\cdot {\bf A}^{<}({\bf r}) \right)
+I_{12}\cos \left({\bf b}_3\cdot {\bf A}^{<}({\bf r})\right) \right]\\
&- \frac{g_b^2}{8} \sum_{n}e^{- G^{>}(0)} \int_{\rv}
   \int_{\bf v} \left[ e^{+ G_{nn}^{>}({\bf v})}-1\right] \left({\bf
       b}_n\cdot ({\bf v}\cdot \nabla) {\bf A}^{<} \right)^2
+ \mbox{ other irrelevant terms}.
 \end{aligned}
 \end{equation}
 \end{widetext}
 We note that above, we have written $G_{nn}^{>}(0)$ simply as
 $G^{>}(0)$, since it takes the same value for all elementary Burgers
 vectors ${\bf b}_n$, $n=1, 2, 3$. Further simplifications lead to the
 desired form
 \begin{widetext}
 \begin{equation}
   \ln\langle e^{-H_{\text{int}}[{\bf A}^{<}+{\bf A}^{>}]}
   \rangle_0^{>}\approx\int_{\rv} 
\left \{ \sum_{n} [g_b e^{-\frac{1}{2} G^{>}(0)} +  J_1 g_b^2 ]
    \cos \left({\bf b}_n\cdot {\bf A}^{<}\right)
     -\frac{g_b^2}{2}\left[J_2 (\partial_i A_j^{<})^2+J_3 \partial_i A_j^< \partial_j A_i^<+J_3 (\grad \cdot {\bf A}^< )^2 \right] \right \},
 \end{equation}
\end{widetext}
where the coefficients $J_1, J_2$ and $J_3$ are defined as:
\begin{subequations}
\begin{equation}
 \begin{split}
   J_1&=I_{12}=I_{13}=I_{23}\equiv \frac{1}{2}e^{- G^{>}(0)} \int_{\bf v}\left[ e^{- G_{12}^{>}({\bf v})}-1\right]\\&= \pi a^2
   \left[e^{\frac{\overline{K}}{16\pi}} I_0\left(\frac{\overline{K}}{8\pi}\right)
+\left(\frac{\overline{K}}{16\pi}-1\right)
   \right] \delta l,
   \end{split}
   \end{equation}
   \begin{equation}
   \begin{split}
   J_2&= \frac{1}{4}\sum_{n=1}^3 (b_2^n)^2 e^{- G^{>}(0)} \int_{\bf v}
   v_1^2 \left[ e^{+ G_{nn}^{>}({\bf v})}-1\right]\\
=&\frac{\pi a^6}{4} 
\left[
  e^{\frac{\overline{K}}{8\pi}}\left(\frac{3}{2}I_0\left(\frac{\overline{K}}{8\pi}\right)
-\frac{3}{4} I_1\left(\frac{\overline{K}}{8\pi}\right)\right)
+\frac{3}{2}\left(\frac{\overline{K}}{16\pi}-1\right) \right] \delta l,
   \end{split}
   \end{equation}
   \begin{equation}
    \begin{split}
   J_3 &=\frac{1}{4} e^{-G^>(0)}\sum_n b_1^n b_2^n \int_{\bf v} v_1 v_2
   \left( e^{+G_{nn}^>({\bf v})}-1\right) \\
&=\frac{3\pi a^6}{16}e^{\frac{\overline{K}}{8\pi}} I_1\left(\frac{\overline{K}}{8\pi}\right) \delta l.
  \end{split}
\end{equation}
\end{subequations}
Above $I_0(x)$, $I_1(x)$ are modified Bessel functions, we have
dropped higher harmonic operators, and have taken, $\delta l \equiv
\ln b \ll 1$.

The above analysis now allows us to identify the renormalized couplings
$K_R^{-1}, B_{R}$ and $g_{b_R}$, 
\begin{subequations}
\begin{eqnarray}
  K_R^{-1}(b)&=& K^{-1} + J_2 g_b^2 ,\\ B_{R}(b)&=&B+J_3 g_b^2,\\
  g_{b_R} (b)&=&g_b e^{-\frac{1}{2} G^{>}(0)} + J_1 g_b^2,
\end{eqnarray}
\end{subequations}
obtained to second-order in $g_b$. The corresponding RG differential
flow equations for the dimensionless couplings
$\overline{K}^{-1}(l)=K^{-1}/a^2$, $\overline{B}(l)=B/a^2$,
$\overline{g}_b(l)=g_b a^2$ are then given by
 \begin{subequations}
 \begin{equation}
  \frac{d \overline{K}^{-1}(l)}{dl}= \frac{3\pi}{8} \left[e^{\frac{\overline{K}}{8\pi}}\left(I_0\left(\frac{\overline{K}}{8\pi}\right)-\frac{1}{2} I_1\left(\frac{\overline{K}}{8\pi}\right)\right) + （\frac{\overline{K}}{16\pi}-1）\right] \overline{g}_b^2,
 \end{equation} 
 \begin{equation} 
  \frac{d\overline{B}(l)}{dl}=\frac{3\pi}{16} e^{\frac{\overline{K}}{8\pi}} I_1\left(\frac{\overline{K}}{8\pi}\right) \overline{g}_b^2,
  \end{equation} 
  \begin{equation}
  \frac{d \overline{g}_b(l)} {dl}=\left(2-\frac{\overline{K}}{8\pi}\right) \overline{g}_b +\pi\left[e^{\frac{\overline{K}}{16\pi}} I_0\left(\frac{\overline{K}}{8\pi}\right)+\left(\frac{\overline{K}}{16\pi}-1\right)\right] \overline{g}_b^2.
\end{equation}
\end{subequations}
Near the melting critical point, $\overline{K}^{-1}(l\to
\infty)=\frac{1}{16\pi}$, $\overline{g}_b(l\to \infty)=0$, we define
the reduced temperature, $x(l)=\frac{16\pi}{\overline{K}}-1$, and
fugacity, $y(l)=e^{-E_{cs}}=\frac{1}{2} \overline{g}_b$. Near the
melting point, their flow equations are given by
\begin{subequations}
\begin{equation}
  \frac{dx(l)}{dl}= 12\pi^2 e^2\left(2I_0(2)- I_1(2)\right) y^2
  \equiv 12 \pi^2 c_1 y^2,
 \end{equation} 
 \begin{equation} 
  \frac{d y(l)} {dl}= 2xy +2\pi e I_0(2)y^2 \equiv 2xy+2\pi c_2 y^2 ,
\end{equation}
\end{subequations}
where $c_1=e^2\left(2I_0(2)- I_1(2)\right) =21.937...$, and $c_2=e
I_0(2)=6.1965...$ are numerical constants, consistent with Halperin
and Nelson\cite{Nelson 79}. 

Following a standard analysis\cite{Kosterlitz 72,Nelson 79}, the
characteristic correlation length $\xi_{xtal-hex}$ near the critical
point at $T\to T_m^-$ can be extracted from above RG flows, giving
\begin{equation}
\xi_{xtal-hex}(T) \sim a e^{-c/|T-T_m|^{\overline{\nu}}},
\end{equation}
with the hexagonal lattice exponent given by
$\overline{\nu}=0.3696\ldots$ and $c$ a nonuniversal constant.

Using the expressions of $K$ and $B$ in terms of the dimensionless Lam\'e elastic constants
$\overline{\mu}=\mu a^2$ and $\overline{\lambda}=\lambda a^2$,
\begin{subequations}
\begin{eqnarray}
\overline{K}^{-1}&=&\frac{1}{4}\left(\frac{1}{\overline{\mu}}+\frac{1}{\overline{\mu}+\overline{\lambda}} \right),\\
\overline{B}&=&\frac{1}{4} \left(\frac{1}{\overline{\mu}}-\frac{1}{\overline{\mu}+\overline{\lambda}} \right),
\end{eqnarray}
\end{subequations}
the RG flow equations for the inverse shear modulus, $\mu^{-1}(l)$ and
the inverse bulk modulus, $[\mu (l) +\lambda(l)]^{-1}$, near the
critical point are then given by,
\begin{subequations}
\begin{eqnarray}
  \frac{d\overline{\mu}^{-1}(l)}{dl}&=& 3\pi e^2 I_0(2) y^2(l),\\
  \frac{d [\overline{\mu}(l)+\overline{\lambda}(l)]^{-1}} {dl}&=& 3\pi  e^2 \left[I_0(2)-I_1(2)\right] y^2(l).\;\;\;\;
\end{eqnarray}
\end{subequations}

\section{Elasticity of 2d crystal subject to an external stress field}
\label{appendixB}
As the crystal is subject to an external stress tensor field $\sigma^e_{ij} ({\bf r})$, we need to add an external term into the elastic energy functional
\begin{equation}
    \mathcal{H}=\frac{1}{2} C_{ij,kl} u_{ij}u_{kl}-\sigma^e_{ij} u_{ij}.
\end{equation}

In the presence of topological defects, the distortion field ${\bf
  u}({\bf r})$ is not single-valued. It and the associated strain
tensor can be decomposed into the single-valued elastic phonon 
and the singular part,
\begin{eqnarray}
u_i&=&\tilde{u}_i+ u_i^s,\\
u_{ij}&=&\tilde{u}_{ij}+ u_{ij}^s.
\end{eqnarray}
To include these topological and phonon degrees of freedom we focus on
the partition function (taking $k_B T = 1$, i.e., measuring coupling
constants in units of thermal energy),
\begin{equation}
\begin{split}
  Z&= \int [d{\bf u}] e^{-\int_{\rv} \mathcal{H}[{\bf u}]}\\
  &=\int [d{\bf u}] \int [d\sigma_{ij}]e^{-\int_{\rv} \mathcal{H}[{\bf u}, \sigma_{ij}]},
\end{split}
\end{equation}
where the trace over ${\bf u}(\rv)$ in the partition function (as in
summing/integrating over the degrees of freedom of the theory)
implicitly includes both phonons and topological defects by allowing
nonsingle-valued distortions. In the second form, above, we decoupled
the elastic energy by introducing a Hubbard-Stratonovich tensor field
-- the {\em symmetric} stress tensor $\sigma_{ij}(\rv)$, with the
resulting Hamiltonian density given by
\begin{equation}
\begin{split}
 \mathcal{H}[{\bf u},\sigma_{ij}]&=\frac{1}{2} C^{-1}_{ij,kl} \sigma_{ij} \sigma_{kl}+i\sigma_{ij}u_{ij}- \sigma^e_{ij} u_{ij}\\&=\frac{1}{2} C^{-1}_{ij,kl} \sigma_{ij}\sigma_{kl}+i\left( \sigma_{ij}+i\sigma^e_{ij}\right) \left( \partial_i \tilde{u}_j + u_{ij}^s \right),
\end{split}
\end{equation}
and $C^{-1}_{ij,kl}=-\frac{\lambda}{4\mu (\mu+\lambda)} \delta_{ij} \delta_{kl}+\frac{1}{2\mu} \delta_{ik}\delta_{jl}$ for a 2D hexagonal lattice.

Tracing over the single-valued phonons $\tilde{\bf u}$, enforces the
divergenceless stress constraint
\begin{equation}
 \partial_i \left(\sigma_{ij}+ i\sigma^e_{ij}\right)=0,
\end{equation}
solved with a scalar Airy stress potential, $\phi(\rv)$,
\begin{equation}
    \sigma_{ij}=\epsilon_{ik}\epsilon_{jl} \partial_k\partial_l\phi  -i\sigma^e_{ij}.
\end{equation}
Expressing the Hamiltonian density in terms of $\phi(\rv)$, and 
integrating by parts in the second linear term, we utilize the defects
conditions, Eqs. (\ref{defn_s}) and (\ref{defn_b}),
\begin{subequations}
\begin{eqnarray}
\epsilon_{ik} \epsilon_{jl} \partial_l \partial_k u_{ij}^s&=&
\epsilon_{ik} \epsilon_{jl} \partial_l \partial_k(\partial_i u_j^s
-\epsilon_{ij}\theta^s),\\
&=&\epsilon_{ki}\partial_k b_i(\rv) +
\epsilon_{ki}\partial_k\partial_i\theta(\rv),\\
&=&\zh\cdot\grad\times{\bf b} + \frac{2\pi}{6}s(\rv),
\end{eqnarray}
\end{subequations}
to obtain
\begin{widetext}
\begin{equation}
\begin{split}
    \mathcal{H}[\phi]&=\frac{1}{2} \tilde{C}_{ij,kl}^{-1} \left(\partial_i\partial_j \phi-i \epsilon_{ia}\epsilon_{jb} \sigma^e_{ij} \right) \left( \partial_k\partial_l \phi-i\epsilon_{kc} \epsilon_{ld} \sigma^e_{kl}\right)+i \frac{1}{2}\epsilon_{ik} \epsilon_{jl} \partial_k \partial_l \phi  \left(\partial_i u_j^s+\partial_j u_i^s \right)\\&=\frac{1}{2} \tilde{C}_{ij,kl}^{-1} \partial_i\partial_j \phi \partial_k\partial_l \phi +i \frac{1}{2} \epsilon_{ik} \epsilon_{jl} \partial_k \partial_l \phi (\partial_i u_j^s -\partial_j u_i^s)+i \epsilon_{ik} \epsilon_{jl} \partial_k \partial_l \phi \partial_j u_i^s-\frac{1}{2}C_{ij,kl}^{-1} \sigma^e_{ij} \sigma^e_{kl}-i \tilde{\tilde C}_{ij,kl}^{-1} \partial_i\partial_j \phi \sigma^e_{kl}\\&=\frac{1}{2} \tilde{C}_{ij,kl}^{-1} \partial_i\partial_j \phi \partial_k\partial_l \phi +i \phi \left( 2\pi s + \hat{\bf z} \cdot \grad \times {\bf b}\right)-\frac{1}{2}C_{ij,kl}^{-1} \sigma^e_{ij} \sigma^e_{kl}-i \tilde{\tilde C}_{ij,kl}^{-1} \partial_i\partial_j \phi \sigma^e_{kl}\\&\equiv \mathcal{H}^{\text{int}}[\phi]+\mathcal{H}^{\text{ext}}[\phi],
\end{split}
\end{equation}
\end{widetext}
where $\tilde{C}_{ij,kl}=\epsilon_{ia} \epsilon_{jb} \epsilon_{kc} \epsilon_{ld} C_{ab,cd}$, $\tilde{\tilde{C}}_{ij,kl}=\epsilon_{ia} \epsilon_{jb}C_{ab,kl}$, and we have used integration by parts. The total elastic energy functional is therefore composed of an internal part and an external part, with the internal part \begin{equation}
\begin{split}
    \mathcal{H}^{\text{int}}=&\frac{1}{2} \tilde{C}_{ij,kl}^{-1} \partial_i\partial_j \phi \partial_k\partial_l \phi +i \phi \left( 2\pi s + \hat{\bf z} \cdot \grad \times {\bf b}\right)\\&+E_{c_s}s^2+ E_{c_b} b^2,
\end{split}
\end{equation}
where we have added the dislocation and disclination core energies
$E_{cb}$ and $E_{cs}$ to account for their short-scales, and the
external part
\begin{equation}
    \mathcal{H}^{\text{ext}}=-\frac{1}{2}C_{ij,kl}^{-1} \sigma^e_{ij} \sigma^e_{kl}-i \tilde{\tilde C}_{ij,kl}^{-1} \partial_i\partial_j \phi \sigma^e_{kl}.
\end{equation}

Alternatively, we can also start by formulating the elastic energy in
terms of both the full strain tensor $\partial_i u_j$ and the bond
angle $\theta ({\bf r})$,
\begin{equation}
\begin{split}
    \mathcal{H}=&\frac{1}{2} C_{ij,kl} (\partial_i u_j-\theta \epsilon_{ij}) (\partial_k u_l-\theta \epsilon_{kl})+\frac{1}{2} K (\partial_i \theta)^2 \\&-\sigma^e_{ij} \left(\partial_i u_j- \theta \epsilon_{ij} \right),
\end{split}
\end{equation}
and get the same result following the procedure of Sec. IIC.

\end{document}